\documentclass[reprint, amsmath,amssymb,aps,pra]{revtex4-1}
\usepackage{subfigure}
\usepackage{graphicx}
\usepackage{dcolumn}
\usepackage{bm}
\usepackage{float}

\begin{document}
\title{Non-Destructive Discrimination of arbitrary set of orthogonal quantum states by NMR using Quantum Phase Estimation}
\email[This paper is dedicated to the memory of Jharana Rani Samal.]{}
\author{V. S. Manu} \email[manuvs@physics.iisc.ernet.in]{} \author{Anil Kumar}\email[anilnmr@physics.iisc.ernet.in]{}
\affiliation{Centre for Quantum Information and Quantum Computing, Department of Physics and NMR Research Centre, Indian Institute of Science,  Bangalore-560012}
\begin{abstract}
An algorithm based on quantum phase estimation, which discriminates 
 quantum states non-destructively within a set of arbitrary orthogonal states, is described and experimentally verified by a NMR quantum information processor.
The procedure is scalable and can be applied to any set  of orthogonal states. Scalability is demonstrated through Matlab simulation. 
\end{abstract}
\maketitle 
\section{Introduction} \label{sec:intro}
Quantum Computing has generated re-newed interest in theory and practice of Quantum Mechanics.
This is largely due to the fact that quantum computers can solve certain problems much faster than classical computers \citep{dj, shor, grover}.
Many efforts are being made to realize a scalable quantum computer using techniques such as  
trapped ions, optical lattices, diamond-based quantum computers, Bose-Einstein condensate based quantum computers, cavity quantum electrodynamics (CQED)
and nuclear magnetic resonance \citep{brennen, jones, nielson,  schmidt, nizovtsev}. 
NMR has become an important experimental tool for demonstrating quantum algorithms, simulating quantum systems, and for verifying various tenets of  
quantum mechanics \citep{jones2, cory2, linden, chuang, lieven, kavita,  ranabir, laflamme, avik2, peng, aram, du}. 

There are  several theoretical protocols available for orthogonal state discrimination \citep{walgate, ghosh, virmani, chen}. Walgate \textit{et al.} showed that, using local operations and classical 
communication (LOCC) multipartite orthogonal states can be distinguished perfectly \citep{walgate}. However if only a single copy is provided and only LOCC is allowed, it cannot discriminate quantum states 
either deterministically or 
probabilistically \citep{ghosh}. Estimation of the phase plays an important role in quantum information processing and is a key subroutine of many quantum algorithms.
When the phase estimation is combined with other quantum algorithms, it can be employed to perform certain computational tasks 
such as quantum counting, order finding and factorization \citep{shor, brassard}. Phase Estimation Algorithm has also been utilised in a recent important application in which the 
ground state of the Hydrogen molecule has been obtained upto 45 bit accuracy in NMR and upto 20 bit accuracy in photonic systems \citep{lanyon}.

By defining an operator with preferred eigen-values, phase estimation can be used logically for discrimination of quantum states with  certainty \citep{nielson}. It preserves the 
state since local operations on ancilla qubit measurements do not affect the state. In this paper we describe an algorithm for non-destructive state discrimination using only phase estimation alone.
The algorithm described in this paper is scalable and can be used for discriminating any set of orthogonal states (entangled or non-entangled). Earlier non-destructive Bell state discrimination has been
described by Gupta \textit{et al.} \citep{panigrahi} and verified experimentally in 
our laboratory by Jharana \textit{et al.} \citep{jharana}. Bell states are specific example of orthogonal entangled states. 
The circuit used for Bell state discrimination \citep{panigrahi} is based  on parity and  phase estimation and 
will not be able to discriminate a superposition state which has no definite parity. For example consider a state $|\psi\rangle=\tfrac{1}{\sqrt{2}}(|00\rangle + |01\rangle)$, 
which belongs to a set of orthogonal states. Here $|00\rangle$ has parity $0$ and $|01\rangle$ has parity $1$.  Hence the above $|\psi\rangle$ does not have a definite parity and cannot be distinguished 
from its other members of the set, by the method of  Gupta \textit{et al.} \citep{panigrahi}.
Sec.\ref{sec:theory} of this paper describes the design of a circuit for non-destructive state discrimination  using phase estimation. Sec. \ref{sec:theory} also contains non-destructive discrimination of 
special cases such as Bell states and three qubit $GHZ$ states using phase estimation. 
Sec.\ref{sec:expt} describes experimental implementation of the algorithm for two qubit states by NMR quantum computer and Sec.\ref{sec:matlab} describes the $Matlab^{\begin{scriptsize}\textregistered\end{scriptsize}}$ 
simulation of non-destructive discrimination of three qubit $GHZ$ states.
\section{Theory}  \label{sec:theory}
For a given eigen-vector $|\phi\rangle$ of a unitary operator $U$,  phase estimation  circuit with $Controlled$-$U$ operator can be used for finding the eigen-value
of $|\phi\rangle$ \citep{nielson}. Conversely the reverse of the algorithm, with defined eigen-values can be used for discriminating eigen-vectors. By logically defining the operators with
preferred eigen-values, the discrimination, as shown here, can be done with certainty.
\subsection{The General Procedure (n-qubit case):}  \label{sec:theory_n}
For $n$ qubit case the Hilbert space dimension is $2^n$, having $2^n$ independent orthogonal states. Hence we need to design a quantum circuit for 
state discrimination for a set  of $2^n$ orthogonal quantum states. 
Consider a set of $2^n$ orthogonal states  $\{\phi_i\}$, where $i=1,2, .... 2^n$.  
The main aim of the discrimination circuit is to make direct correlation between the elements of $\{\phi_i\}$ and  possible product states of ancilla qubits. 
As there are $2^n$ states, we  need $n$ ancilla qubits for proper discrimination. 

\begin{figure*} \label{fig:circuit_nqbt}
\begin{center} 
 \includegraphics[width=10cm,height=3.2cm]{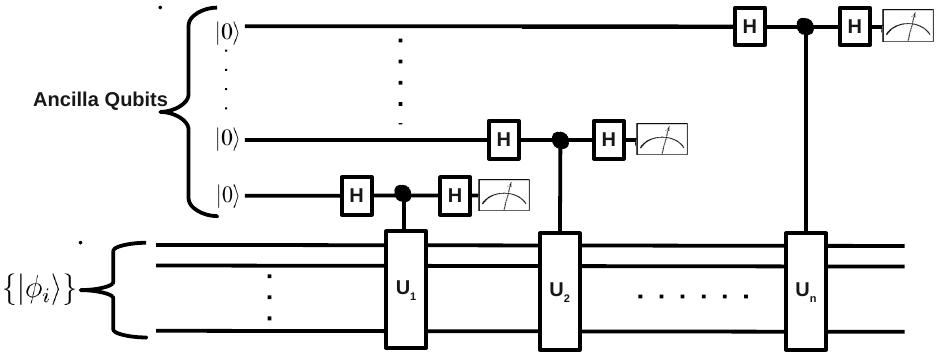}
 \caption{The general circuit for non-destructive Quantum State Discrimination. For discriminating $n$ qubit states it uses $n$ number of
 ancilla qubits with $n$ controlled operations. $n$ ancilla qubits are first prepared in the state $|00...0\rangle$. Here H represents Hadamard 
transform and the meter represents a measurement of the qubit state. The original state encoded in n qubits is preserved(not  destroyed).}.
\end{center}
\end{figure*}

The discrimination circuit requires $n$ Controlled Operations. Selecting these $n$ operators $\{U_j\}$ (where $j=1,2,... n$) is the main task in designing the algorithm. 
The set $\{U_j\}$  depends on the $2^n$ orthogonal states in such a way that the set of orthogonal vectors forms the eigen-vector set of the operators, 
with eigen-values $\pm1$. The sequence of $+1$ and $-1$ in the eigen-values should be defined in a special way, as outlined below.
 Let $\{e_j^i\}$ (with $i=1,2... 2^n$) be the eigen-value array of $U_j$, and it should satisfy following conditions. \\
\textit{Condition \#1}: Eigen-value arrays $\{e_{j}^i\}$ of all operators $\{U_j\}$ should contain equal number of +1 and -1, \\
\textit{Condition \#2}: For the first operator $U_1$, the eigen-value array $\{e_1^i\}$ can be any possible sequence of +1 and -1 with \textit{Condition \#1}, \\
\textit{Condition \#3}: The restriction on eigen-value arrays starts from $U_{j=2}$ onwards. The eigen-value array  ($\{e_2^i\}$) of operator $U_2$ should not be equal to $\{e_1^i\}$ or its complement,
while still satisfying the \textit{Condition \#1}. \\
\textit{Condition \#4}: By generalizing the \textit{Condition \#3}, the eigen-value array  ($\{e_k^i\}$) of operator $U_k$ should not be equal to $\{e_m^i\}$ $(m=1,2,...k-1)$ or its complement. \\

Let $M_j$ be the diagonal matrix formed by eigen-value array  $\{e_j^i\}$ of $U_j$. The operator $U_j$ is directly related to $M_j$ by a  unitary transformation given by,
\begin{equation}
U_j=V^{-1} \times M_j \times  V,
\end{equation}
where V is the matrix formed by the column vectors  $\{|\phi_i\rangle\}$,
V = [ $|\phi_1\rangle$ ~~  $|\phi_2\rangle$ ~~  $|\phi_3\rangle$ .....  $|\phi_n\rangle$].

The circuit diagram for implementation of Phase Estimation Algorithm (PEA) to discriminate orthogonal states using the $Controlled$-$U_j$ operations 
such that the original state is preserved for further use in any qauntum circuit  is  shown in Fig.1.

As  the  eigen-values defined are either $+1$ or $-1$, the final ancilla qubit states will be in product state (without superposition), and hence can be measured with certainty.
It can be  shown that the selection of specific operator set $\{U_j\}$ with the conditions discussed above makes direct correlation between $2^n$ product states of  
ancilla qubit and elements of $\{|\phi_i\rangle\}$ so that ancilla measurements can discriminate the state. 
\subsection{Single qubit case:} \label{sec:theory_1}
For a single qubit system, the Hilbert space dimension is 2. So we can discriminate a state from a set of two orthogonal states.
Consider an illustrative example with the orthonormal set as \{ $|\phi_1\rangle = \tfrac{1}{\sqrt{2}}(|0\rangle + |1\rangle)$, 					
$|\phi_2 \rangle= \tfrac{1}{\sqrt{2}}(|0\rangle - |1\rangle)$ \}. The quantum circuit for this particular case can be designed 
by following the general procedure discussed in Sec.\ref{sec:theory_n}.
The V matrix for the given states $\{|\phi_1\rangle, |\phi_2\rangle\}$ is,
\begin{center}
V= $\dfrac{1}{\sqrt{2}}\begin{pmatrix}
 1 & 1 \\
 1 & -1
\end{pmatrix}. $
\end{center}
According to the rules given in Sec.\ref{sec:theory_n}, $M$ can be either 
$\begin{pmatrix}
 1 & 0 \\
 0 & -1
\end{pmatrix}  $ or $\begin{pmatrix}
 -1 & 0 \\
 0 & 1
\end{pmatrix}.  $ ~~~~~~~~~~~
For $M$ = $\begin{pmatrix}
 1 & 0 \\
 0 & -1
\end{pmatrix}  $,
\begin{equation}
~~~~~~~~U=V^{-1}\times M \times V = \begin{pmatrix}
 0 & 1 \\
 1 & 0
\end{pmatrix} .
\end{equation}

The circuit diagram for this case is identical to Fig.1, having only one work and one ancilla qubit. It can be easily shown that, the ancilla qubit measurements are directly correlated with the input states. For the selected $M_1$, if the given 
state is $|\phi_1\rangle$  then ancilla will be in the state $|0\rangle$  and if the given state is $|\phi_2\rangle$  ancilla will be in the state $|1\rangle$.

For a general set  \{$|\phi_1\rangle =(\alpha|0\rangle + \beta|1\rangle)$, $|\phi_2 \rangle= (\beta|0\rangle -\alpha |1\rangle)$\} (where $\alpha$  and $\beta$ are real numbers satisfying, 
$|\alpha|^2+ |\beta|^2 = 1$), operator $U$ for eigenvalue array $\{1, -1\}$ can be shown as, \\

\begin{equation}
U=\begin{pmatrix}
 Cos(\theta) & Sin(\theta) \\
 Sin(\theta) &-Cos(\theta)
\end{pmatrix},
\end{equation} \\   with  $\theta = 2 \times Tan^{-1}(\dfrac{\beta}{\alpha})$.
\subsection{Two qubit case:} \label{sec:theory_2}
The Hilbert space dimension of two qubit system of is four. Consider an illustrative example with a set of orthogonal states 
\begin{align}  \notag
\{|S(\alpha,\beta)\rangle\}=\{ (\alpha|00\rangle + \beta|01\rangle), (\alpha|10\rangle + \beta|11\rangle), \\ 
 (\beta|10\rangle - \alpha|11\rangle), (\beta|00\rangle - \alpha|01\rangle)\},
\end{align}

where $\alpha$  and $\beta$ are real numbers satisfying, $|\alpha|^2+ |\beta|^2 = 1$. 
This set is so chosen that the states are (a)orthogonal, (b)not entangled,  (c)different from Bell states, (d)do not have definite parity and (e)contain 
single-superposed-qubits (SSQB) (in this case second qubit is superposed).
Using the general procedure discussed above, we can select the eigen-value arrays for two operators $U_1$ and  $U_2$ as
\begin{equation}
\{e_1\}=\{1,1,-1,-1\}, ~~~~~~~~~~~\{e_2\}=\{1,-1,1,-1\}.
\end{equation}
$U_1$ and  $U_2$, the unitary transformation of the diagonal matrices formed by $\{e_1\}$ and $\{e_2\}$ are, \\

\begin{equation}
U_1 = \begin{pmatrix}
 Cos(\theta)  &  Sin(\theta)  & 0 &  0 \\
 Sin(\theta)  & -Cos(\theta)  & 0 &  0 \\
 0  &  0  &  Cos(\theta) &  Sin(\theta)\\
 0  &  0  &  Sin(\theta) & -Cos(\theta)\\
\end{pmatrix},
\end{equation}
\begin{equation}
U_2 = \begin{pmatrix}
 Cos(\theta)  & Sin(\theta)   & 0 &  0\\
 Sin(\theta)  & -Cos(\theta)  & 0 &  0\\
 0  & 0  & -Cos(\theta) & -Sin(\theta)\\
 0  &  0 & -Sin(\theta) &  Cos(\theta)\\
\end{pmatrix},
\end{equation}
where,   $\theta = 2 \times Tan^{-1}(\dfrac{\beta}{\alpha})$. \\


\begin{table} 
\begin{center}
\begin{tabular}{|c|c|c|}
\hline
states & Ancilla-1  & Ancilla-2 \\
\hline
$|\phi_1\rangle$ & $|0\rangle$ & $|0\rangle$ \\
$|\phi_2\rangle$ & $|0\rangle$ & $|1\rangle$ \\
$|\phi_3\rangle$ & $|1\rangle$ & $|0\rangle$ \\
$|\phi_4\rangle$ & $|1\rangle$ & $|1\rangle$ \\
\hline
\end{tabular}
\label{table1}
\caption{State of ancilla qubits for different input states for two qubit orthogonal states.} 
\end{center}
\end{table}

The output state of the  ancilla qubit run through all possible product states as input state changes, as listed in  Table.I. The
quantum circuit for two qubit state discrimination is shown in Fig.2a.

\subsubsection{Special case ($\alpha=\beta=\frac{1}{\sqrt{2}}$):} \label{sec:theory_spcl}
The set of orthogonal states are, 
\begin{align}  \notag
\{|S(\tfrac{1}{\sqrt{2}},\tfrac{1}{\sqrt{2}})\rangle\}=  \{|\phi_i\rangle\} 
= \{\tfrac{1}{\sqrt{2}}(|00\rangle + |01\rangle),\\\tfrac{1}{\sqrt{2}}(|10\rangle + |11\rangle),
\tfrac{1}{\sqrt{2}}(|10\rangle - |11\rangle),\tfrac{1}{\sqrt{2}}(|00\rangle - |01\rangle)\}.
\end{align}

The operators $U_1$ and $U_2$ can be found by substituting the value of $\theta = \tfrac{\pi}{2}$ in (5) and (6),
\begin{equation}
U_1 = \dfrac{1}{\sqrt{2}}
 \begin{pmatrix}
 0  & 1  & 0 &  0\\
 1  & 0  & 0 &  0\\
 0  & 0  & 0 &  1\\
 0  &  0 & 1 &  0\\
\end{pmatrix}   
and  ~~
U_2 = \dfrac{1}{\sqrt{2}}
 \begin{pmatrix}
 0  & 1  & 0 &  0\\
 1  & 0  & 0 &  0\\
 0  & 0  & 0 & -1\\
 0  &  0 &-1 &  0\\
\end{pmatrix}.
\end{equation}
The quantum circuit for the set (Eqn.8) is same as the general case of any set of two qubit orthogonal states(Fig.2a).
Experimental implementation of this case has been performed using NMR and is described in  Sec. \ref{sec:expt}.
\subsection{Bell state discrimination:} \label{sec:theory_Bell}
Bell states are maximally entangled two qubit states (also known as Einstein-Podolsky-Rosen states) \citep{epr}. They play a crucial role in several applications of quantum computation 
and quantum information theory. They have been used for teleportation, dense coding and entanglement swapping \citep{bennet1, bennet2, pan, zukowski}. Bell states have also found application in remote state preparation, where 
a known state is prepared in a distant laboratory \citep{pati}. Hence, it is of general interest to distinguish Bell states without disturbing them.
The complete set of Bell states is,
\begin{align} \notag
\{|B_i\rangle\} = \{\tfrac{1}{\sqrt{2}}(|0_20_1\rangle + |1_21_1\rangle),\tfrac{1}{\sqrt{2}}(|0_20_1\rangle - |1_21_1\rangle), \\
\tfrac{1}{\sqrt{2}}(|0_21_1\rangle + |1_20_1\rangle),\tfrac{1}{\sqrt{2}}(|0_21_1\rangle - |1_20_1\rangle)\}
\end{align}
Bell states form an orthogonal set. Hence one can design a circuit for Bell state discrimination using only phase estimation.
The circuit diagram is same as that shown in Fig.2a with different $U_1$ and $U_2$. For eigen-value arrays $\{e_1\}=\{1,-1,1,-1\}, \{e_2\}=\{-1,1,1,-1\}$.
$U_1$ and  $U_2$ are obtained as, \\
\begin{equation}
U_1 = \begin{pmatrix}
 0  & 0  & 0 &  1\\
 0  & 0  & 1 &  0\\
 0  & 1  & 0 &  0\\
 1  & 0  & 0 &  0\\
\end{pmatrix}   and~~ U_2 = \begin{pmatrix}
 0  & 0  & 0 & -1\\
 0  & 0  & 1 &  0\\
 0  & 1  & 0 &  0\\
-1  & 0  & 0 &  0\\
\end{pmatrix}.
\end{equation}
The controlled operators ($C-U_1$ and $C-U_2$) for phase estimation which involves 3-qubit operators  can be written as the product of 2-qubit operators as,
\begin{align}
C-U_1 & = C-NOT^3_1 \times C-NOT^3_2 , \\   
 C-U_2 & = C-\pi^2_1 \times C-NOT^4_1 \times C-NOT^4_2 \times C-\pi^2_1.  \notag
\end{align}
Here qubits 1 and 2 are work qubits in which the Bell states are encoded and 3 and 4 are the ancilla qubits.
Here $C-NOT^i_j$ represents C-NOT operation with control on $i^{th}$ qubit and target on $j^{th}$ qubit.
The splitting  of three qubit operator into two qubit operators is needed for the implemention of $C-U_1$ and $C-U_2$.

There already exists an algorithm for non-destructive discrimination of Bell state by Gupta \textit{et al.} \citep{panigrahi}, which has also been experimentally implemented in NMR by Jharana \textit{et al.} \citep{jharana}. 
The circuit of Gupta \textit{et al.} \citep{panigrahi}  is based on parity and phase measurement and will fail  for a superposed state which has no definite parity. 
However for non-destructive discrimination of Bell states using the present phase estimation algorithm is similiar to Gupta's circuit where the parity estimation is replaced by 
modified phase estimation.

\subsection{$GHZ$ state discrimination:} \label{sec:theory_ghz}
$GHZ$ states are maximally entangled multi qubit states \citep{ghz}. GHZ states have been used in several quantum algorithms such as 
quantum secret sharing, controlled dense coding and quantum key distribution \citep{hao, xiao, ying}.
These algorithms make use of entanglement and hence it is important to discriminate $GHZ$ states by preserving their entanglement.

All $n$ qubit $GHZ$ states form an orthogonal set (without definite parity). Hence a circuit can be designed for discriminating general $n$ qubit $GHZ$ states using only phase estimation. 

Consider the case of three qubit $GHZ$ states,  which are 
\begin{align}
\notag
\{|G_i\rangle\} = \{  \tfrac{1}{\sqrt{2}}(|000\rangle + |111\rangle), \tfrac{1}{\sqrt{2}}(|000\rangle - |111\rangle),\\ \notag
\tfrac{1}{\sqrt{2}}(|001\rangle + |110\rangle), \tfrac{1}{\sqrt{2}}(|001\rangle - |110\rangle),\\ \notag
\tfrac{1}{\sqrt{2}}(|010\rangle + |101\rangle), \tfrac{1}{\sqrt{2}}(|010\rangle - |101\rangle), \\ 
\tfrac{1}{\sqrt{2}}(|011\rangle + |100\rangle), \tfrac{1}{\sqrt{2}}(|011\rangle - |100\rangle)\} 
\end{align}

Here we need three ancilla qubits and have to implement three controlled operators for state discrimination. Verification of the NMR experiment to discriminate 
such states has been 
carried out here using $Matlab^{\begin{scriptsize}\textregistered\end{scriptsize}}$ and the parameters of a four qubit NMR system, as described in Sec.\ref{sec:matlab}.
\section{Experimental Implementation by NMR}  \label{sec:expt}
\subsection{Non-Destructive Discrimination of two qubit orthogonal states:} \label{sec:expt_2}
Experimental implementation of the quantum state discrimination(QSD) algorithm has been performed here for 2 qubit case for the orthogonal set (Eqn.7).

\begin{figure*}  \label{fig:circuit_2qbt_splt}
\begin{center}
 \includegraphics[width=8.9cm,height=4.3cm]{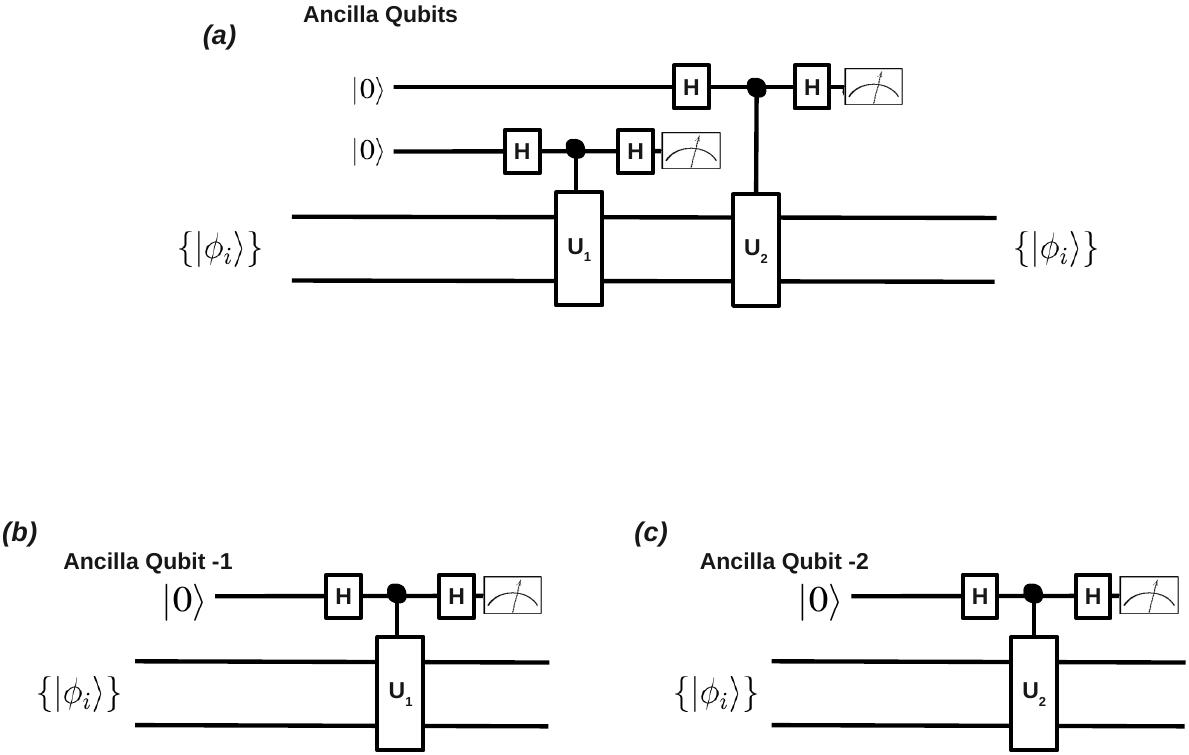}
 \caption{(a) Two qubit State discrimination circuit for Experimental implementation in three qubit NMR quantum computer, (b) and (c) are splitting of the circuit-(a)
 into two circuits with single ancilla measurements.}
\end{center}
\end{figure*}

The discrimination circuit diagram shown in Fig.2a, needs a 4 qubit system. As two ancilla qubits are independent of each other, 
following \citep{jharana} one can split the experiment into 2 measurements with a single ancilla qubit (Fig.2b and 2c).

The NMR implementation of the discrimination algorithm starts with (i) preparation of the pseudo-pure state followed by (ii) creation of input state 
(iii) Quantum Phase estimation with operators $\{U_j\}$. Finally the measurement on ancilla qubits yields the result.

The experiment has been carried out at 300K in $11.7 T$ field in a Bruker $AV500$ spectrometer using a triple resonance QXI probe.
The system chosen for the implementation of the discrimination algorithm is Carbon-13 labeled Dibromo-fluoro methane $^{13}CHFBr_2$, where $^1H$, $^{19}F$ and $^{13}C$ act as the three qubits\citep{vanedrsypen, avik3}.
The $^1H$, $^{19}F$ and $^{13}C$ resonance frequencies at this field are 500, 470 and 125 MHz, respectively. The scalar couplings between the spins are: 
$J_{HC}= 224.5 Hz$, $J_{HF} = 49.7 Hz $ and $J_{FC} = - 310.9 Hz$(Fig.3).

The NMR Hamiltonian for a three qubit weakly coupled spin system is \citep{ernst},
\begin{align}
H = \displaystyle\sum\limits_{i=0}^3\nu_iI_z^i+ \displaystyle\sum\limits_{i<j=1}^3J_{ij}I_z^iI_z^j,
\end{align}

where $\nu_i$ are the Larmor frequencies and the $J_{ij}$ are the scalar couplings. The starting point of any algorithm in an 
NMR quantum information processor is the equilibrium density matrix, 
which under high temperature and high field approximation is in a highly mixed state represented by\citep{avik3},
\begin{align} \notag
\rho_{eq} ~~~~\propto ~~~~\gamma_H I_z^H + \gamma_C I_z^C +\gamma_F I_z^F  \\
= \gamma_H(I_z^H +0.94I_z^F +0.25I_z^C).
\end{align}
There are several methods for creating pseudo pure states (PPS) in NMR from equlibrium state \citep{cory1, gershenfeld, cory2, sallt}. We have utilized the spatial averaging technique \citep{cory2} for creating 
pseudo pure states as described in \citep{avik3}. 
The spectra for equlibrium and $|000\rangle$ PPS are shown in Fig.3.
\begin{figure*}  \label{fig:spctras_eq_pps_new}
\begin{tabular}{cc}

\subfigure{\includegraphics[width=4.5cm,height=2.7cm]{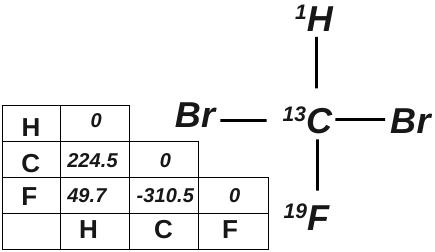}}
\subfigure{\includegraphics[width=10cm,height=3cm]{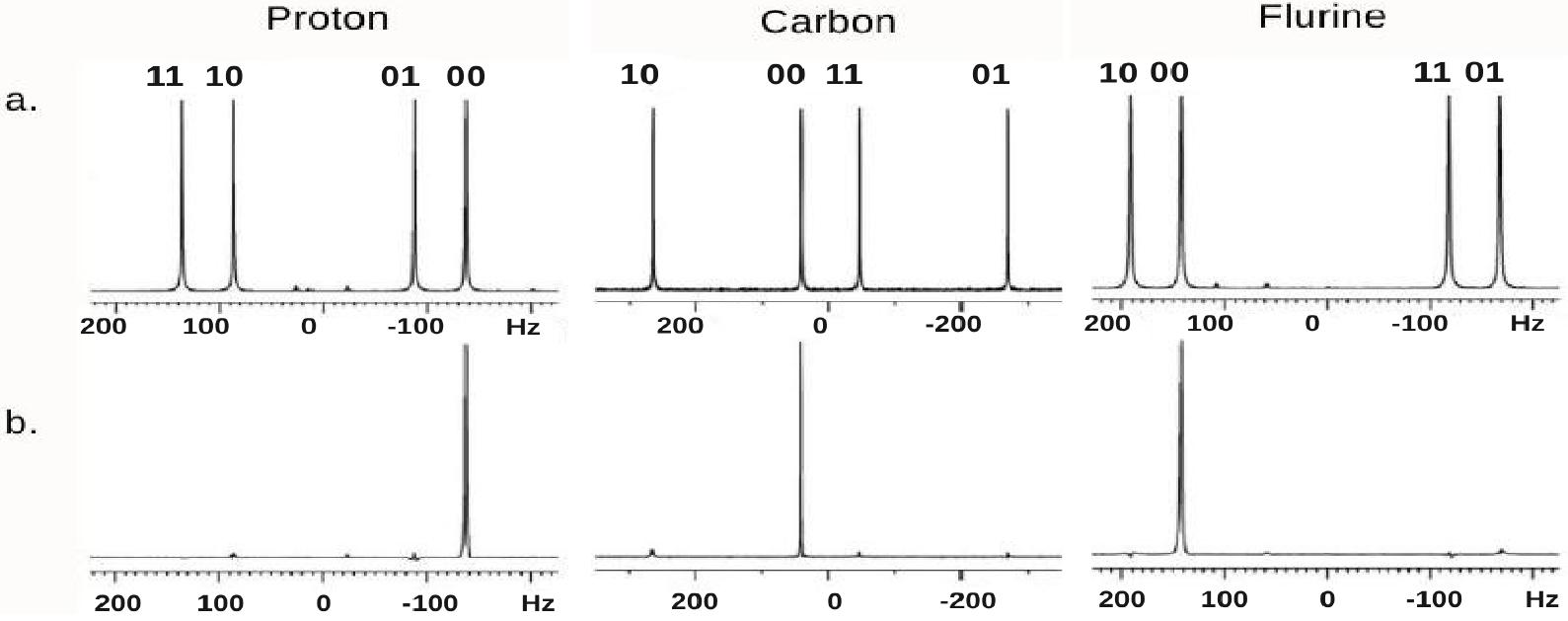}}
\end{tabular}~\\
\caption{The three qubit NMR sample used for experimental implementation. The nuclear spins $^1H$, $^{19}F$ and $^{13}C$  are 
used as the three qubits. (a) Equlibrium Spectra of proton, carbon and fluorine, (b) Spectra corresponds to the created $|000\rangle$ pseudo pure state.
These spectra are obtain by using $90^o$ measuring pulse on each spin.}
\end{figure*}

For Phase Estimation algorithm, due to its high sensitivity, proton spin has been utilized as the ancilla qubit; and  the two qubit states, 
to be discriminated, are encoded in carbon and fluorine spins. 			
As the measurements are performed only on ancilla qubit, we record only proton spectra for non-destructive discrimination of the state of carbon and fluorine.
The state of the ancilla qubit can be identified by the relative phase of the spectra. We set the phase such that a positive peak indicates that the proton 
was initially in state $|0\rangle$.
\subsubsection*{Implementation of $Controlled$-$U_1$ and $U_2$:}
For the set of orthogonal states given in eqn.(9) the $U_1$ and $U_2$ are given in Eqn.(8). Let  $H_1$ and  $H_2$ be the effective Hamiltonians 
for $Controlled$-$U_1$ and $Controlled$-$U_2$  propagators such that,  \\
$Controlled$-$U_1=exp(i H_1)$, \\
$Controlled$-$U_2=exp(i H_2)$.\\ 
where $H_1$ and $H_2$, in terms of product operators \citep{ernst} are obtained as, \\
$H_1=(\dfrac{\pi}{4} I - \dfrac{\pi}{2} I^1_z - \dfrac{\pi}{2} I^3_x  + \pi I^1_z I^3_x)$,  \\
$H_2=(\dfrac{\pi}{4} I - \dfrac{\pi}{2} I^1_z- \pi I^2_z I^3_x  + 2 \pi I^1_z I^2_z I^3_x)$.\\
Since the various terms in $H_1$ and $H_2$ commute with each other, one can write,

\begin{align}\notag 
Controlled-U_1  = exp(i H_1) & \\\notag 
 =  exp(i(\dfrac{\pi}{4} I - \dfrac{\pi}{2} I^1_z -& \dfrac{\pi}{2} I^3_x  + \pi I^1_z I^3_x))    \\\notag 
=  exp(i\dfrac{\pi}{4} I)   \times exp( -i \dfrac{\pi}{2} I^1_z) \times &  exp(-i \dfrac{\pi}{2} I^3_x) \times   exp(i\pi I^1_z I^3_x), \\\notag 
Controlled-U_2   =exp(i H_2) &\\ \notag 
= exp(i(\dfrac{\pi}{4} I - \dfrac{\pi}{2} I^1_z -  &\pi I^2_z I^3_x  + 2 \pi I^1_z I^2_z I^3_x)) \\ \notag
= exp(i\dfrac{\pi}{4} I)\times   exp(-i&\dfrac{\pi}{2} I^1_z) \times  exp(-i\pi I^2_z I^3_x)\\ 
&~~~ \times exp(i2\pi I^1_z I^2_z I^3_x).
\end{align}

As the decomposed terms commute with each other, these propagators can be easily implemented in NMR(Fig.4). Single spin operators such as $I_x$, $I_y$ are implemented using R.F pulses. 
The $I_z$ operator is implemented using composite $z$ rotation pulses in NMR ($(\frac{\pi}{2})_{-x}(\frac{\pi}{2})_y(\frac{\pi}{2})_x$) \citep{ml,sorensen}. 
Two spin product terms such as  $I^i_z I^j_x$ are implemented using scalar coupling Hamiltonian evolution sandwiched between two $(\frac{\pi}{2})_y$ pulse on $j$ spin \citep{avik3}.
The three spin product operator terms are implemented using cascades of two spin operator evolutions(\textit{Tseng et al.} \citep{tseng}).

\begin{figure*}  \label{fig:u1u2}
\begin{tabular}{cc}
\centering
\subfigure[$Controlled$-$U_1$]{\includegraphics[width=4cm,height=3cm]{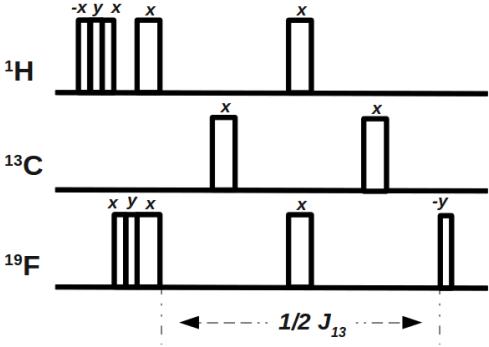}}  
\subfigure[$Controlled$-$U_2$]{\includegraphics[width=10cm,height=3cm]{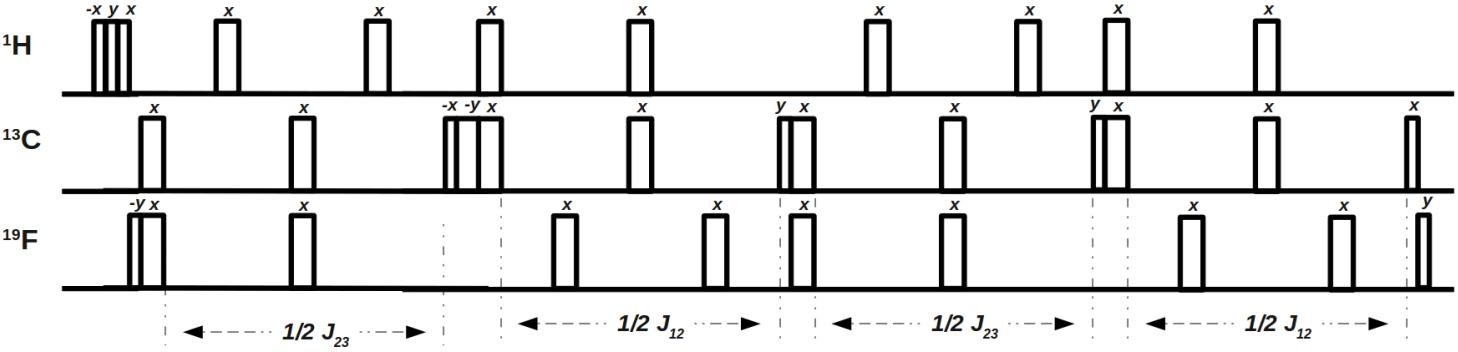}}  
\end{tabular}
\caption{The pulse sequence for $Controlled$-$U_j$ operators for two qubit orthogonal states shown in (14) (Here narrow pulses indicate ($\tfrac{\pi}{2}$) pulses  and broad pulses 
indicate $\pi$ pulses with the phase given above the pulse).}
\end{figure*}

\begin{figure*}  \label{fig:finalspectra}
\includegraphics[width=15cm,height=3.5cm]{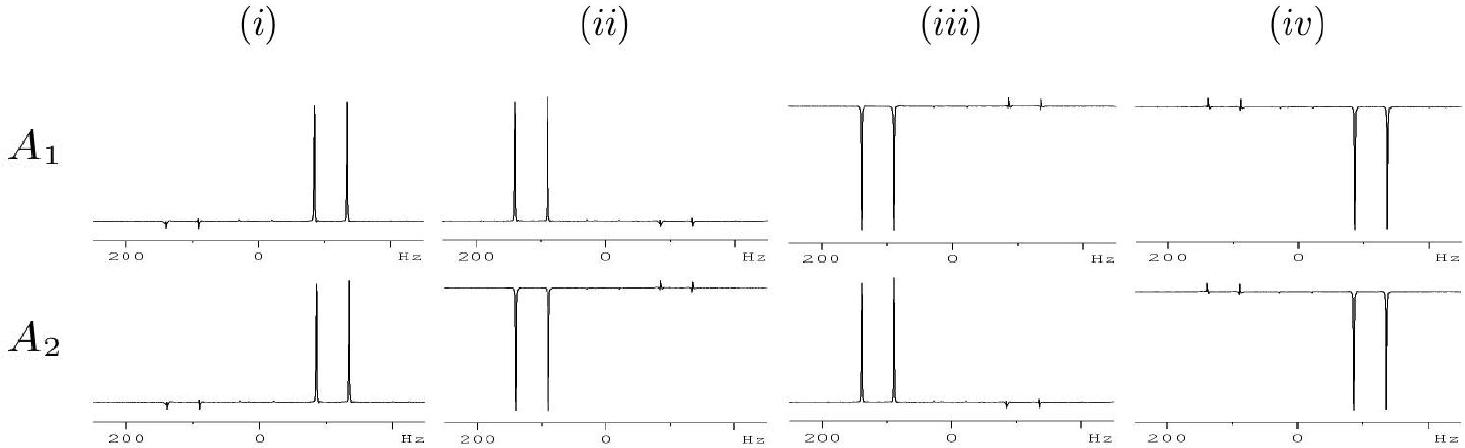} 
\caption{Ancilla (proton spin) spectra of final state for two qubit state discrimination algorithm for $(i) ~~ |\phi_1\rangle$, $(ii) ~~ |\phi_2\rangle$, $(iii) ~~ |\phi_3\rangle$,
 $(iv) ~~ |\phi_4\rangle$ states. $A_1$ and $A_2$ are results of two measurements on single ancilla(Fig.4b and 4c respectively)  qubit-1 and 2 (here it is two experiments with same ancilla qubit). 
These spectra are obtained with a $90^o$ measuring pulse on the ancilla(proton) qubit at the end of the pulse sequence. }
\end{figure*}
The experimental results are shown in Fig.5. Proton spectra shows the state of ancilla qubit, which in-turn can be used  for 
discrimination of two qubit state in carbon and fluorine spins. Positive peaks in Fig. 5 means ancilla is in qubit state $|0\rangle$ and negative peak indicates 
ancilla qubit is in state $|1\rangle$. Thus spectra in Fig.5 indicates that $(i),(ii),(iii),(iv)$ are respectively $|\phi_1\rangle,|\phi_2\rangle,|\phi_3\rangle, |\phi_4\rangle$ (Table.I).
To compute fidelity of the experiment, complete density matrix tomography has been carried out (Fig.6).

\begin{figure*}  \label{fig:tomo}
\includegraphics[width=14cm,height=16.3cm]{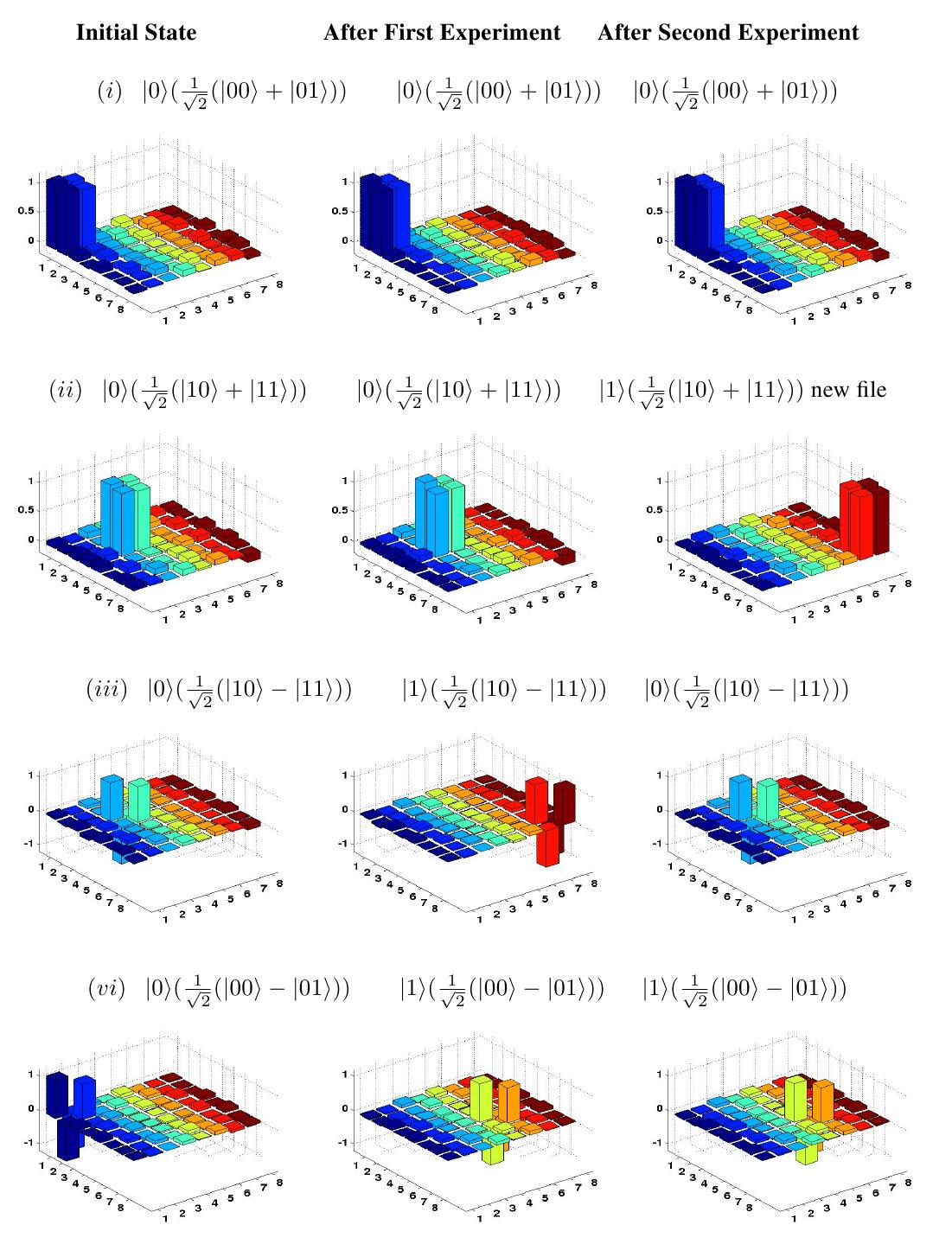}
\caption{Density Matrix Tomography of the initial and final states of QSD circuit.  First qubit is the ancilla. It is evident that the state of $2^{nd}$ and $3^{rd}$ qubits are preserved.
(here $1 \rightarrow |000\rangle$, $2 \rightarrow |001\rangle$, $3 \rightarrow |010\rangle$, $4 \rightarrow |011\rangle$, $5 \rightarrow |100\rangle$, $6 \rightarrow |101\rangle$, 
 $7 \rightarrow |110\rangle$, $8 \rightarrow |111\rangle$.)}
\end{figure*}
The experimental results are in agreement with the Table.I with an \textquoteleft average absolute deviation\textquoteright ~\citep{avik} of 4.0\% and \textquoteleft maximum absolute deviation\textquoteright ~\citep{avik} of 7.2\%,
providing the desired discrimination.
\section{Three qubit $GHZ$ state Discrimination using $Matlab^{\textregistered}$ Simulation:} \label{sec:matlab}
Non-destructive discrimination of  the three qubit maximally entangled ($GHZ$) states using only Phase Estimation algorithm as described in Sec.\ref{sec:theory} 
in NMR has also been performed using a $Matlab^{\begin{scriptsize}\textregistered\end{scriptsize}}$ simulation. This 
simulation verifies the principle involved but does not include any decoherence  or pulse imperfection effects. 
The three qubit GHZ states form a set $\{G_i\}$ given by eqn.(13) can be re-expressed as,
\begin{align}
\{|G_i\rangle\} =  \{ |\phi_1\rangle, |\phi_1\rangle, |\phi_3\rangle,........ |\phi_8\rangle \}
\end{align}
The discrimination of a 3-qubit $GHZ$ state using phase estimation requires 3 work qubits and 3 ancilla. We divide the 6 qubit quantum circuit into three circuits. 
Each circuit has three work qubits and a single ancilla. There are several possibilities for eigen-value sets which will satisfy the sets of conditions discussed in Sec.\ref{sec:theory}. 
Consider one such set, 
\begin{align}
\notag
\{e_1\}&=\{1,-1,1,-1,1,-1,1,-1\}, \\
\{e_2\}&=\{-1,1,1,-1,1,-1,-1,1\}, \\\notag
\{e_3\}&=\{-1,1,1,-1,-1,1,1,-1\}. 
\end{align}
For this eigen-value set (18), the $Controlled-U_j$ operators can be written as 
\begin{align}
\notag
Controlled-U_1 = C-NOT^a_1 &\times C-NOT^a_2 \times  C-NOT^a_3,  \\ \notag
Controlled-U_2 = C-\pi^2_3  \times C&-NOT^a_1 \times C-NOT^a_2 \\ \notag
& \times  C-NOT^a_3 \times C-  \pi^2_3,      \\ \notag
Controlled-U_3  = C-\pi^1_3  \times C&-NOT^a_1 \times C-NOT^a_2 \\ 
\times  C-NOT&^a_3    \times C-  \pi^1_3.
\end{align}

\begin{table}   
\begin{center}
\begin{tabular}{|c|c|c|c|}
\hline
state & Measurement-1  & Measurement-2 & Measurement-3 \\
\hline
$|\phi_1\rangle$ & $|0\rangle$ & $|1\rangle$ & $|1\rangle$\\
$|\phi_2\rangle$ & $|1\rangle$ & $|0\rangle$ & $|0\rangle$\\
$|\phi_3\rangle$ & $|0\rangle$ & $|0\rangle$ & $|0\rangle$\\
$|\phi_4\rangle$ & $|1\rangle$ & $|1\rangle$ & $|1\rangle$\\
$|\phi_5\rangle$ & $|0\rangle$ & $|0\rangle$ & $|1\rangle$\\
$|\phi_6\rangle$ & $|1\rangle$ & $|1\rangle$ & $|0\rangle$\\
$|\phi_7\rangle$ & $|0\rangle$ & $|1\rangle$ & $|0\rangle$\\
$|\phi_8\rangle$ & $|1\rangle$ & $|0\rangle$ & $|1\rangle$\\
\hline
\end{tabular}
\label{table1}
\caption{State of ancilla qubits for different input states of Eqn.(13) and (17).} 
\end{center}
\end{table}

Splitting of four qubit operator into two qubit operators is needed for its experimental implementation (Fig.7)
Here 1,2 and 3 are the work qubits, in which the $GHZ$ state is encoded and \textquoteleft a\textquoteright ~ is the ancilla qubit.
The results of ancilla qubit measurements are tabulated in Table.II.
~\\
\begin{figure*}  \label{fig:u1u2u3}
\centering
\includegraphics[width=21cm,height=13cm,angle=90]{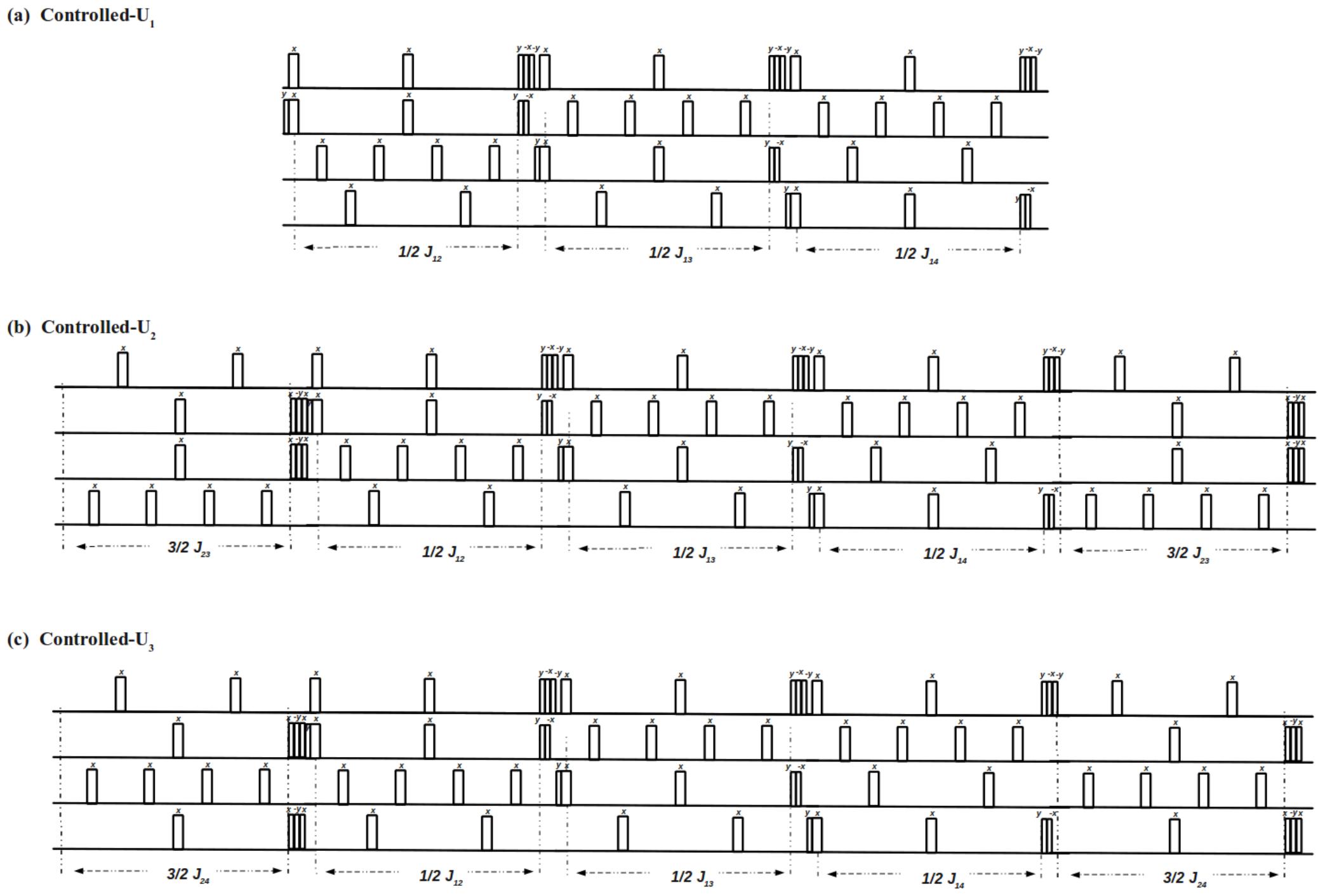}
\caption{Pulse sequence for Controlled operators in $GHZ$ state discrimination. (Here narrow pulses indicate ($\tfrac{\pi}{2}$) pulses  and broad pulses 
indicate $\pi$ pulses with the phase given above the pulse).}
\end{figure*}

NMR simulation has been carried out using the parameters of a well known 4-qubit system, crotonic acid with all carbons labelled by $^{13}C$(Fig.8) \citep{laflamme}. 
The density matrix tomography of the 
$Matlab^{\begin{scriptsize}\textregistered\end{scriptsize}}$ experiment for a few selected($|\phi_1\rangle $, $|\phi_4\rangle$ and $|\phi_7\rangle$)
 $GHZ$ states are shown in Fig.10. This confirms that the method of Phase Estimation discussed in Sec.\ref{sec:theory} can be used 
for discrimination of $GHZ$ states without destroying them.

\begin{figure*}[h]  \label{fig:crotonic}
\begin{center}
\includegraphics[width=6cm,height=3.7cm]{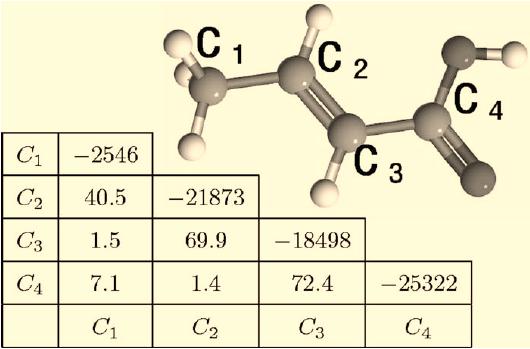}
\end{center}
\caption{The chemical structure, the chemical shifts and spin-spin coupling of a $^{13}$C labelled Crotonic Acid. The four $^{13}C$ spins act as four qubits \citep{laflamme}.}
\end{figure*}

\begin{figure*}  \label{fig:ghztomo}
\begin{tabular}{cccc}
~~ & First Experiment & Second Experiment & Third Experiment \\~\\
$(i)$ & $|0\rangle_a (\tfrac{1}{\sqrt{2}}(|000\rangle+|111\rangle))$ & $|1\rangle_a (\tfrac{1}{\sqrt{2}}(|000\rangle+|111\rangle))$ & $|1\rangle_a (\tfrac{1}{\sqrt{2}}(|000\rangle+|111\rangle))$ \\
~ &\includegraphics[width=4.2cm,height=4cm]{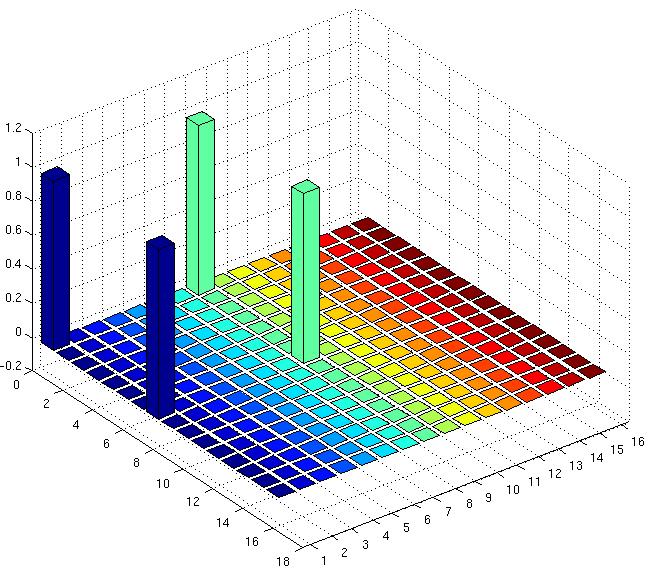} & \includegraphics[width=4.2cm,height=4cm]{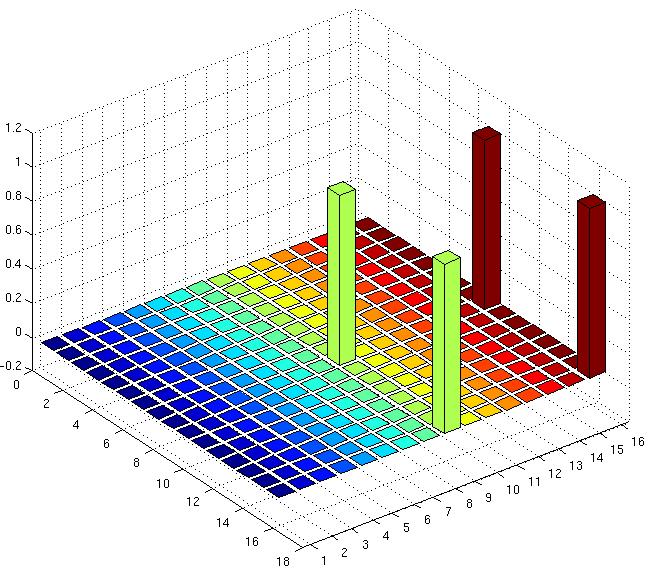}  & \includegraphics[width=4.2cm,height=4cm]{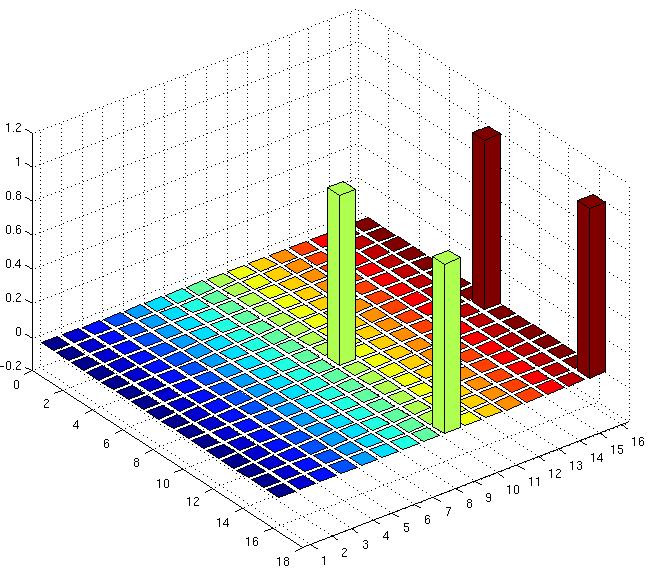} \\~\\~\\
$(ii)$ & $|1\rangle_a (\tfrac{1}{\sqrt{2}}(|001\rangle-|110\rangle))$ & $|1\rangle_a (\tfrac{1}{\sqrt{2}}(|001\rangle-|110\rangle))$ & $|1\rangle_a (\tfrac{1}{\sqrt{2}}(|001\rangle-|110\rangle))$ \\
~ & \includegraphics[width=4.2cm,height=4cm]{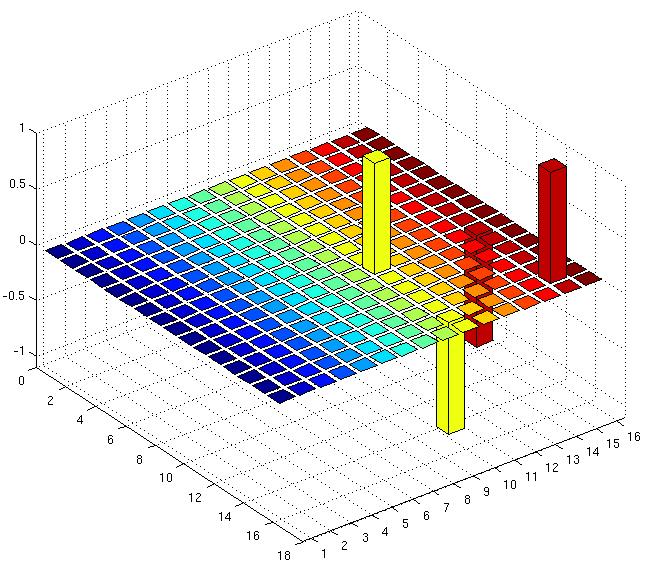} & \includegraphics[width=4.2cm,height=4cm]{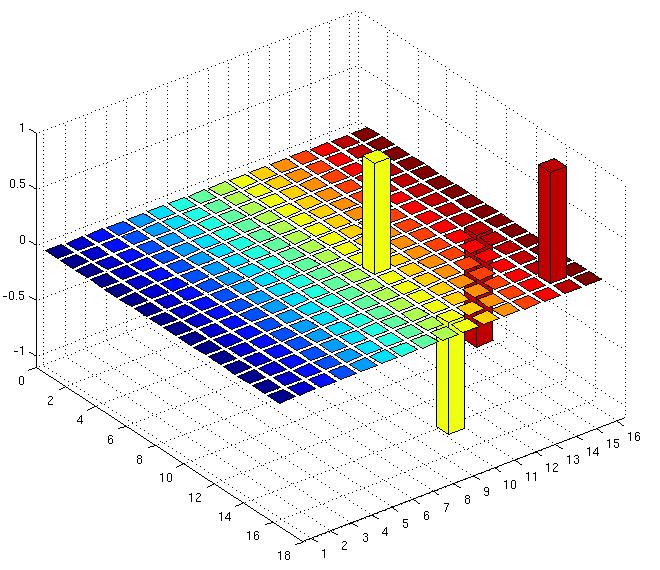}  & \includegraphics[width=4.2cm,height=4cm]{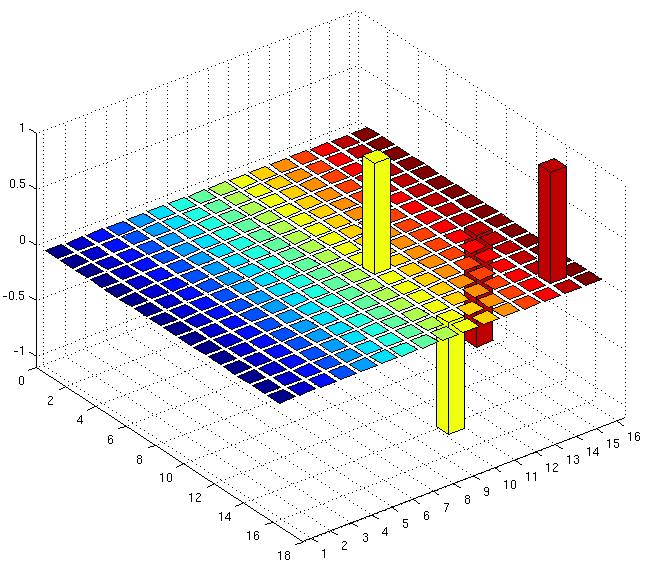} \\~\\~\\
\end{tabular}
\caption{ $Matlab^{\textregistered}$  simulation results for $GHZ$ state discrimination. The simulated spectras are shown for three  $GHZ$ states $|\phi_1\rangle $ and $|\phi_4\rangle$.
It is evident from final density matrix that the $GHZ$ states are preserved.
(Here $1 \rightarrow |0000\rangle$, $2 \rightarrow |0001\rangle$, $3 \rightarrow |0010\rangle$, $4 \rightarrow |0011\rangle$, $5 \rightarrow |0100\rangle$, $6 \rightarrow |0101\rangle$, 
 $7 \rightarrow |0110\rangle$, $8 \rightarrow |0111\rangle$,  $9 \rightarrow |1000\rangle$, $10 \rightarrow |1001\rangle$, $11 \rightarrow |1010\rangle$, $12 \rightarrow |1011\rangle$, $13 \rightarrow 
|1100\rangle$, $14 \rightarrow |1101\rangle$, $15 \rightarrow |1110\rangle$, $16 \rightarrow |1111\rangle$. First qubit is the ancilla)}
\end{figure*}

\section*{Conclusion}
A general scalable method for non-destructive quantum state discrimination of a set of orthogonal states using quantum phase estimation algorithm has been 
descibed, and experimently implemented for a  two qubit case by NMR. As the direct measurements  are performed only on the 
ancilla, the discriminated states are preserved. The generalization of the algorithm is 
illustrated by discrimination of $GHZ$ states using a $Matlab^{\begin{scriptsize}\textregistered\end{scriptsize}}$ simulation.
~\\~\\
\bibliographystyle{unsrtnat}
\bibliography{stdisb}
\end{document}